\documentclass[sigconf]{acmart}

\usepackage{graphicx}
\usepackage{colortbl}
\usepackage{etoolbox}
\usepackage{multirow}
\usepackage{makecell}
\usepackage{color}
\usepackage[utf8]{inputenc}
\usepackage{xspace}
\usepackage{colortbl}
\usepackage{pifont}
\usepackage{tabularx}

\AtBeginDocument{%
  }

\copyrightyear{2025}
\acmYear{2025}
\setcopyright{acmcopyright} 
\acmConference[IDC '25]{Interaction Design and Children}{June 23--26, 2025}{Reykjavik, Iceland}
\acmBooktitle{Interaction Design and Children (IDC '25), June 23--26, 2025, Reykjavik, Iceland}\acmDOI{10.1145/3713043.3727051}
\acmISBN{979-8-4007-1473-3/2025/06}

\begin{document}
\title{Scratch Copilot: Supporting Youth Creative Coding with AI}

\author{Stefania Druga}
\orcid{0000-0002-5475-8437}
\affiliation{%
  \institution{Google Deep Mind}
  \city{Seattle}
  \state{Washington}
  \country{USA}
   \postcode{98195}}
\email{druga@google.com}

\author{Amy J. Ko}
\orcid{0000-0001-7461-4783}
\affiliation{%
  \institution{University of Washington}
  \city{Seattle}
  \state{Washington}
  \country{USA}
  \postcode{98195}}
\email{ajko@uw.edu}

\renewcommand{\shortauthors}{Druga, et al.}

\begin{teaserfigure}
\centering
\includegraphics[width=\columnwidth]{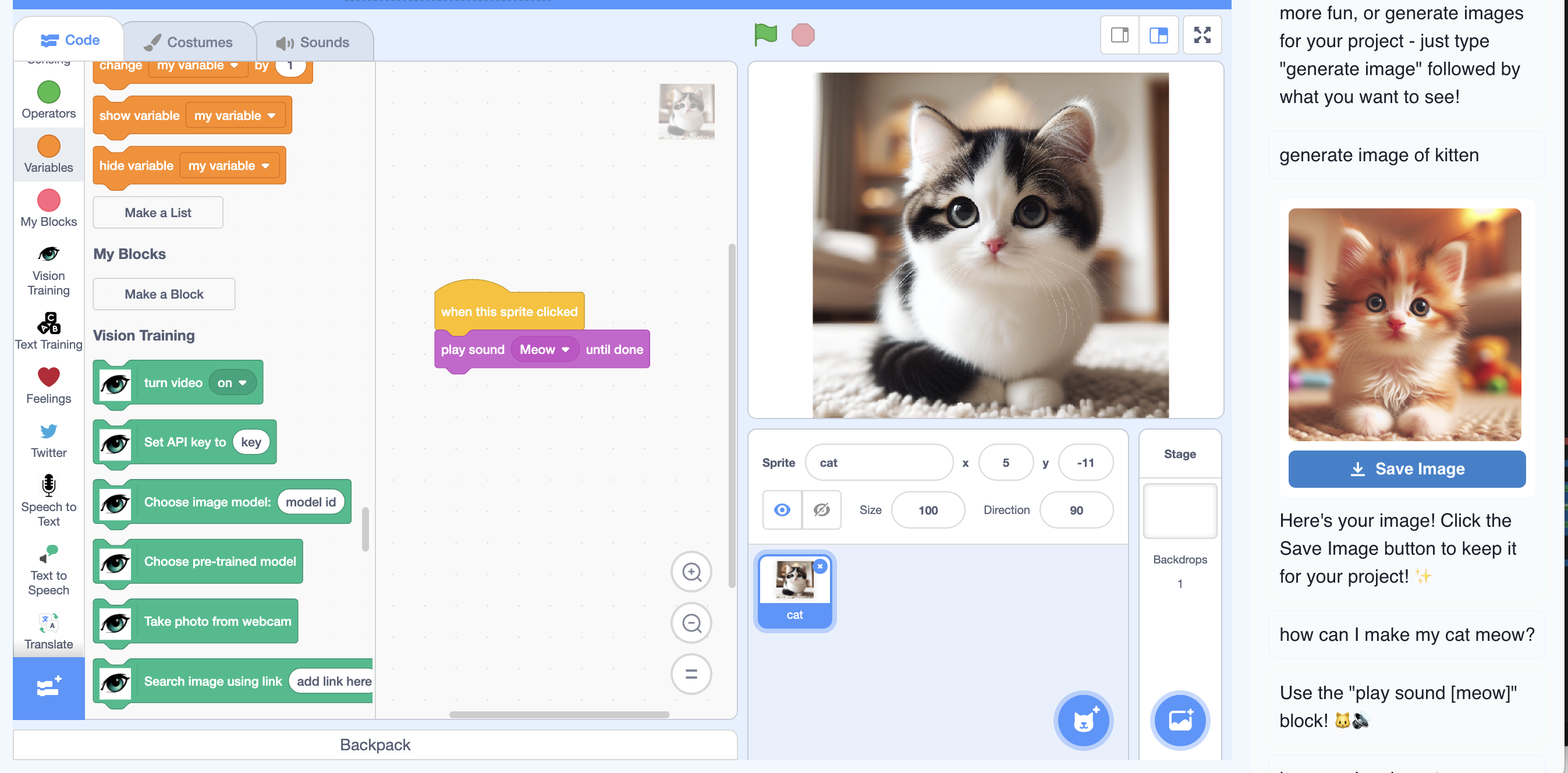}
\caption{Cognimates interface showing coding blocks, AI chat, and image generation features.}
\label{fig:platform_ui}
\end{teaserfigure}

\begin{abstract}
Creative coding platforms like Scratch have democratized programming for children, yet translating imaginative ideas into functional code remains a significant hurdle for many young learners. While AI copilots assist adult programmers, few tools target children in block-based environments. Building on prior research \cite{druga_how_2021,druga2023ai, druga2023scratch}, we present \textbf{Cognimates Scratch Copilot}: an AI-powered assistant integrated into a Scratch-like environment, providing real-time support for ideation, code generation, debugging, and asset creation. This paper details the system architecture and findings from an exploratory qualitative evaluation with 18 international children (ages 7--12). Our analysis reveals how the AI Copilot supported key creative coding processes, particularly aiding ideation and debugging. Crucially, it also highlights how children actively negotiated the use of AI, demonstrating strong agency by adapting or rejecting suggestions to maintain creative control. Interactions surfaced design tensions between providing helpful scaffolding and fostering independent problem-solving, as well as learning opportunities arising from navigating AI limitations and errors. Findings indicate Cognimates Scratch Copilot's potential to enhance creative self-efficacy and engagement. Based on these insights, we propose initial design guidelines for AI coding assistants that prioritize youth agency and critical interaction alongside supportive scaffolding.
\end{abstract}



\maketitle

\section{Introduction}
Creative coding platforms, such as Scratch \cite{maloney2010scratch}, have become instrumental in democratizing computer programming, empowering millions of young learners to express themselves creatively through code. Scratch, with its block-based visual interface, significantly lowers the entry barrier to coding, making it accessible to children as young as 7 years old. However, despite the platform's accessible visual programming interface, novice middle school learners often encounter difficulties in translating their imaginative ideas into functional code and require scaffolding and guidance \cite{jimenez2018usability,zhang2019systematic}. These challenges range from grappling with programming logic to debugging errors and effectively utilizing the platform's features. Consequently, young coders require robust support mechanisms to navigate these hurdles, fully realize their creative potential, and develop crucial computational thinking skills \cite{papert1980mindstorms, wing2006computational}.

The rise of Artificial Intelligence (AI) presents a unique opportunity to address these challenges and enhance the creative coding experience for youth. AI-powered coding assistants have shown promise in supporting adult programmers and university students in various aspects of software development \cite{madaan2023learning, chen2023teaching, sarsa2022automatic}. Tools like GitHub Copilot and Cursor AI demonstrate the potential of AI to boost coding productivity in text-based environments \cite{GovTech2024Copilot, restackAI}. However, a significant gap exists in the development and research of AI-powered coding tools specifically designed for young learners within visual programming contexts. While platforms like MIT RAISE AI Playground \cite{raiseAIPlayground} and eCraft2Learn extensions \cite{ecraft2learn} enable children to \textit{create} AI projects by integrating AI features into block-based coding, they do not provide AI-driven assistance \textit{within} the creative coding process itself. To the best of our knowledge, no AI Copilot tool has been explicitly tailored to support middle-schoolers' creative coding with visual programming languages like Scratch.

We address these limitations through our Cognimates Scratch Copilot Tool\footnote{The Scratch Copilot platform is accessible at: \url{http://cognimatescopilot.com/}}, an AI assistant that:
\begin{itemize}
\item Integrates directly with Scratch's visual programming language through natural language dialogue
\item Employs a question-driven pedagogy to scaffold creative problem-solving
\item Adapts responses to childrens' cultural contexts
\end{itemize}

Cognimates Scratch Copilot is a fully functional AI-powered assistant designed to enhance creative coding experiences for youth on the Cognimates platform. It provides multifaceted support, including ideation assistance, intelligent code generation, debugging guidance, and code explanation, all within the familiar Scratch environment. The development of Cognimates Scratch Copilot was guided by insights from our prior participatory design study with 10 children aged 7-12 and 9 parents from 6 different US states \cite{druga2023aifriendsdesignframework}.

This study details the system architecture of Cognimates Scratch Copilot and presents findings from an evaluation with an international group of 18 children (ages 7--12) from 11 different countries, focusing on how Cognimates Scratch Copilot supports their creative coding processes. Our evaluation employs qualitative methods to assess the impact of the tool on various aspects of creative coding.

Our preliminary findings indicate that participants found Cognimates Scratch Copilot to be a constructive tool, readily adopting it for their creative projects and expressing sustained engagement. The AI Copilot proved to be effective in aiding ideation, code implementation, debugging, and answering open-ended queries. Participants appreciated the copilot's interactive approach, particularly its tendency to ask clarifying questions, fostering a more engaging learning experience.

However, the study also revealed significant variations in children's prior exposure to AI technologies and AI literacy across different geographic regions, highlighting the importance of culturally responsive design \cite{solyst2022insights, moralez2019computing}. Participants also provided valuable feature requests for future platform design iterations.

Our contributions include:
\begin{enumerate}
\item The first tool for AI-assisted visual programming in middle school education
\item Empirical evidence of cross-cultural effectiveness from 18 international participants
\item Ethical design principles for child-AI co-creative programming grounded in constructionist learning theory
\end{enumerate}

Overall, our findings underscore the potential of AI-powered copilots like Cognimates Scratch Copilot to significantly enhance youth creative coding experiences and fill a critical gap in existing AI education tools. We present initial design guidelines for AI-assisted coding tools targeted at young learners, emphasizing user-centered design, scaffolding, and culturally responsive approaches. By fostering creative self-efficacy, boosting engagement, and facilitating computational thinking skills, Cognimates Scratch Copilot can enable the next generation of digital creators.
\section{Related Work}
Our work builds upon existing research in AI-powered coding tools, inclusive AI education for youth, creative coding pedagogies, and ethical design considerations for AI in education. However, a key novelty of our work lies in addressing the gap of AI-powered copilots specifically designed for programming environments for middle school learners, an area underexplored by prior studies.

\subsection{AI-Powered Coding Assistants for Youth: Addressing the Novelty Gap}
Existing AI-powered coding assistants, such as GitHub Copilot \cite{GovTech2024Copilot} and Cursor AI \cite{cursorAI}, have demonstrated significant benefits for professional software development, enhancing productivity and streamlining workflows in text-based coding environments. These tools leverage advanced AI to provide code suggestions, auto-completion, and debugging support, primarily targeting adult programmers and university-level computer science students. Research has validated the effectiveness of LLMs in enhancing code authoring and understanding for these demographics \cite{Kazemitabaar2023Studying, leinonen2023using}. However, these tools are not designed to address the unique needs of younger learners, such as metacognitive scaffolding \cite{loksa2022metacognition}. 

While some platforms like PictoBlox \cite{pictoBlox} and RAISE AI Playground \cite{raiseAIPlayground} integrate AI and machine learning features into block-based environments, they do not offer AI as a copilot to assist youth directly within the coding process. For example, MIT RAISE AI Playground provides extensions for Scratch to enable children to program AI functionalities, but it lacks an embedded AI assistant that guides coding, debugging, and creative ideation within Scratch projects. Similarly, while tools like Teachable Machine \cite{teachableMachine} empower children to build machine learning models, they are separate from visual programming environments designed for broader creative expression. Therefore, Cognimates Scratch Copilot distinguishes itself by directly addressing the gap in AI-powered coding assistance tailored for youth creative coding within a visual programming language, offering a novel approach to supporting young learners in this domain.

\subsection{Culturally Responsive and Inclusive AI Education for Youth}
Creating equitable and effective AI education tools for youth requires careful consideration of cultural responsiveness and inclusivity. Research highlights significant socio-cultural disparities in children's access to and perceptions of AI \cite{Kim2023AttitudesAI, FreireSnchezIntergenerationalDO}. Studies show that students from diverse cultural backgrounds exhibit varying levels of AI literacy and different attitudes towards AI technologies, influenced by factors such as geographic location, socioeconomic status, and prior exposure to technology \cite{druga2019inclusive, Chiu2023AILiteracy}. A meta-review of AI education in the Asia-Pacific region underscores the need for culturally adapted curricula to ensure equitable access to AI literacy for K–12 students \cite{Chiu2023AILiteracy}. 

Furthermore, developmental psychology emphasizes the importance of age-appropriate design in technology for children \cite{NeugnotCerioli2024TheFO}. AI tools for young learners must be designed to scaffold cognitive load, foster creative self-efficacy, and align with children's developmental stages \cite{NeugnotCerioli2024TheFO, Druga2023ScratchCE}. Existing AI education resources often lack specific adaptations for diverse cultural contexts and may not fully address the developmental nuances of middle school learners. In contrast, Cognimates Scratch Copilot is designed with these considerations in mind. 

Our tool design was based on a participatory design study the involved children with a wide range of socio-economic backgrounds and who spoke 5 languages other than English \cite{druga2023aifriendsdesignframework}, informing the development of a tool that aims to be culturally responsive and developmentally appropriate. By providing multilingual support, adapting to different levels of AI literacy, and focusing on fostering creative self-efficacy, Cognimates Scratch Copilot seeks to contribute to more inclusive and equitable AI education for youth globally.

\subsection{Creative Coding Pedagogies and AI Support}
Creative coding pedagogies emphasize a bricolage approach to learning, where programming is viewed as an iterative process of tinkering, experimenting, and improvising \cite{mclean2012computer, turkle1990epistemological}. Platforms like Scratch are designed to support this style of learning, encouraging exploration and playful experimentation \cite{maloney2010scratch}. However, while creative coding environments are inherently engaging, they do not automatically guarantee the development of computational thinking skills \cite{woo2022problem}. 

Novice learners often require scaffolding and guidance to effectively translate their creative ideas into functional code and to develop robust problem-solving strategies within these environments \cite{brennan2015hf}. Prior research on Creative Support Tools (CSTs) highlights the importance of tools that are integrated into creators' daily practices and support various stages of the creative process, including ideation, implementation, and reflection \cite{Frich2019-zj, Kim2017-mosaic}. While some CSTs exist for creative domains like music and visual arts, few are tailored to support creative coding for youth, particularly in visual programming environments. Furthermore, existing AI tools in education often focus on structured learning environments and outcome-oriented tasks, potentially overlooking the more open-ended and exploratory nature of creative coding. 

Cognimates Scratch Copilot aims to bridge this gap by providing AI support that is specifically aligned with creative coding pedagogies. By offering ideation prompts, debugging assistance, and code explanations within the Scratch environment, Cognimates Scratch Copilot seeks to enhance the creative coding process while fostering computational thinking skills in a way that is congruent with constructionist learning theories \cite{resnick2013designing, papert1980mindstorms}. Moreover, by focusing on question-asking and iterative refinement, Cognimates Scratch Copilot encourages an exploratory and iterative approach to learning, rather than simply providing direct answers, thus aligning with constructionist learning principles and the spirit of creative coding.

\subsection{Ethical and Design Considerations for AI Copilots in Education}
As AI tools become increasingly integrated into education, ethical and design considerations are paramount. Current AI ethics frameworks emphasize mitigating harmful biases and ensuring fairness, accountability, and transparency \cite{newman2024want}. However, in the context of AI for youth creative coding, ethical considerations extend beyond bias mitigation to include promoting positive development, fostering agency, and ensuring responsible use. Concerns exist about the potential for over-reliance on AI, which might hinder the development of fundamental problem-solving and critical thinking skills in young learners \cite{shah2024impact}. Furthermore, the ``black box'' nature of some AI systems can make it difficult for children to understand how AI suggestions are generated, potentially undermining trust and sense-making \cite{wang2024large}. While guidelines for ethical AI in education are emerging \cite{newman2024want, han2023aistory}, actionable strategies for designing AI tools that promote positive youth development and agency in creative coding are still needed. Moreover, the potential for AI to influence children's creative self-efficacy and sense of authorship requires careful consideration \cite{newman2024want, joanganzcooneycenterAI}. Cognimates Scratch Copilot is designed with these ethical and design considerations in mind. We prioritize transparency by providing code explanations and encouraging user interaction through question-asking. To foster agency and avoid over-reliance, Cognimates Scratch Copilot requires active user confirmation for AI suggestions and promotes iterative refinement of code. Furthermore, we aim to evaluate the impact of Cognimates Scratch Copilot on children's creative self-efficacy and to develop design guidelines that promote responsible and ethical use of AI in youth creative coding education. By addressing these ethical and design challenges, Cognimates Scratch Copilot seeks to contribute to the responsible and beneficial integration of AI into creative learning environments for young people.

\section{Method}
Our study addressed the question: \textit{How might we support children to engage in collaborative creative coding with an AI Copilot?} To answer this, we developed an AI-assisted coding platform for youth, building on our prior empirical work that identified priority needs in this space while promoting creative self-efficacy \cite{druga2023aifriendsdesignframework}.

\subsection{Selection and Participation of Children}
We recruited 18 children aged 7-12 from diverse backgrounds (11 countries: USA, Spain, Singapore, China, Mexico, Romania, Jamaica, Canada, India, Israel, New Zealand). This 7-12 age range was selected to capture perspectives across late elementary and early middle school, a key developmental period for introducing creative coding concepts and exploring interactions with emerging technologies like AI. Children had a variety of experiences with AI tools and programming, which was captured through intake questionnaires. The study occurred via video conference, with participants grouped by age and prior experience to enable better analysis of demographic interactions with the AI Copilot. All parents and children older than age 7 signed consent forms. The first author explained the study details to participants before obtaining consent. All sessions were conducted individually between one researcher (the first author) and one child participant. The University of Washington Institutional Review Board reviewed and approved the study protocol.

\subsection{Study Procedure}
Each study session, including the pre-coding, AI-enhanced coding, and reflection phases, lasted approximately 40 to 50 minutes.

\textbf{Phase 1: Pre-AI Coding}. Participants began by discussing their prior Scratch experience, then created a simple project (e.g., making a sprite say ``Hello'') without AI assistance. This established baseline coding skills and familiarity with the platform interface (Figure \ref{fig:platform_ui}).

\textbf{Phase 2: AI-Enhanced Coding}. After completing initial tasks, participants chose to either modify their projects using the AI Copilot or start new ones. During this phase, participants typically focused on developing a single project, allowing for deeper engagement with the AI Copilot's features within that context. Researchers suggested AI use when participants asked coding or platform-related questions (e.g., ``How do I make my cat meow?''). Our goal was to understand how children might engage with the AI Copilot. We did not force them to use it, but rather encouraged use of the assistant when it might help them get unblocked, to observe their interaction with it in a semi-authentic collaborative context. If the AI provided unhelpful responses after two attempts, we offered direct assistance. This researcher scaffolding occurred in approximately 10 instances where help was sought, often to clarify AI suggestions or guide platform navigation, particularly for younger participants or those entirely new to Scratch.

\textbf{Phase 3: Post-AI Reflection}. Participants reflected on their AI experiences through semi-structured interviews covering:
\begin{itemize}
\item AI's role in their creative process (likes/dislikes)
\item Perceptions of AI capabilities and limitations
\item Ideas for improving AI collaboration
\item Broader views on AI's future in coding education
\end{itemize}
All sessions concluded with participants retaining platform access, as many preferred continuing AI use rather than stopping abruptly.

\textbf{Systematic Study Scaffolding.} To accommodate varying participant experience levels, we implemented the scaffolding strategy described above. The first author (who conducted all sessions) adopted a structured approach: when participants asked coding or platform questions, they first were encouraged to use the AI Copilot. If AI was unsuccessful after two attempts, researcher support followed. This was framed as an invitation to explore the capabilities of the AI Copilot and as an option to try another modality of support when needed, rather than as an explicit instruction to use the AI.

This scaffolding prioritized AI as first responder to:
\begin{itemize}
\item Encourage AI capability exploration aligned with research goals
\item Maintain consistent support across sessions
\item Minimize researcher bias in problem-solving
\end{itemize}
This approach of providing researcher support and scaffolding was motivated by the exploratory nature of our research. Our primary goal was to investigate the potential of AI Copilot for creative coding, as opposed to conducting a summative evaluation of the prototype.

\subsection{AI Copilot Platform}
\begin{figure*}[thpb]
\centering
\includegraphics[width=5.2in]{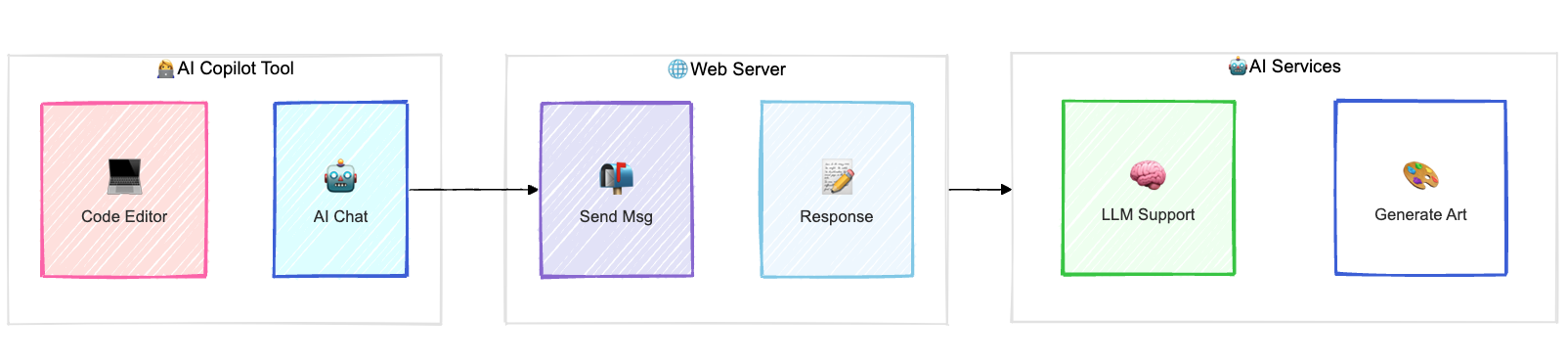}
\caption{Cognimates Scratch Copilot System Architecture}

\label{fig:system_architecture}
\end{figure*}

The AI Copilot Tool, depicted in Figure \ref{fig:system_architecture}, was developed as a web-based platform integrating a visual coding environment with an integrated AI chat assistant. The client-side of the platform is built using React, a JavaScript library for building user interfaces, and Chakra UI, a component library providing a set of accessible and composable building blocks. The AI Copilot directly interfaces with the OpenAI API, leveraging the capabilities of GPT-4o for text-based chat and DALL-E 3 for image generation.

The platform presents two main interactive areas: a coding area and an AI chat area (see Figure \ref{fig:platform_ui}). The \textit{Coding Platform} includes a \textit{Coding Blocks Library} and a \textit{Coding Area}. The \textit{Coding Blocks Library} features a collection of pre-built visual blocks representing various programming actions, similar to the popular Scratch programming environment. The \textit{Coding Area} serves as an interactive canvas where users can arrange and connect these visual blocks to create a functioning program controlling a sprite, visualizing the program's output in real-time.

The \textit{AI Chat} component provides a user interface for interacting with the AI Copilot.  It includes a text-based chat window where users can enter natural language prompts to seek assistance, explore ideas, or request images. These prompts are processed by the GPT-4o model. Markdown rendering is used to display rich text and code snippets. To generate images, users can enter ``generate image of'' followed by a description of the desired image. This triggers DALL-E 3, which generates an image based on the prompt, allowing the user to preview the image and download the assets directly. The chat window is positioned next to the coding area to create an integrated workflow, and facilitate a fluid interaction with the AI Copilot.

To ensure safe and reliable use, the AI Copilot integrates several safety mechanisms. Content filtering through OpenAI’s built-in safety system protects against prompts containing inappropriate or harmful content. By ``inappropriate or harmful'' we mean content that is not youth appropriate, promotes violence, involves hate speech, or reveals personally identifiable information about other individuals. To make the user experience more accessible, the platform delivers kid-friendly error messages when the AI is unable to process a request. A ``kid-friendly'' interface is one where text is simplified and actions are easily discoverable. It prioritizes ease of use and avoids jargon, ensuring children easily understand and navigate the content. Furthermore, rate limiting handling and secure API key management mechanisms are in place to guarantee a high level of security and prevent misuse.

The workflow of using the AI Copilot Tool is as follows:
\begin{enumerate}
    \item Users create visual programs in the Scratch coding area.
    \item When users require assistance, they type a prompt in the AI Chat area.
     \item User prompts are transmitted to the back-end server.
    \item The AI Copilot, based on the user's message, can perform one of these operations:
    	\begin{itemize}
            \item Provide coding help and answers
            \item Generate images
     \end{itemize}
     \item Responses from the AI Copilot, including images, are returned to the application and displayed in the chat area.
     \item Users can see their conversation history in the chat window and interact with the AI Copilot as needed.
\end{enumerate}

The system architecture, as illustrated in Figure \ref{fig:system_architecture}, consists of a client-side application, a server, and AI services. On the user's computer, a React application renders the entire platform within a web browser, including components for chat, image display, and the Scratch coding interface. The \textit{Web Server} is built using Express.js, which handles incoming user messages through defined routes and manages the logic to communicate with the AI services. Finally, the \textit{AI Services} include OpenAI's GPT-4o for text-based responses and DALL-E 3 for generating images.

\subsection{AI Copilot Persona}

The AI Copilot’s persona was designed to be a supportive peer and helpful coding assistant, drawing upon prior research on social robots and educational agents to create a positive and engaging learning environment. The \textit{system prompt} was carefully crafted to encourage guiding questions and positive feedback rather than direct answers, following our prior co-design study \cite{druga2023aifriendsdesignframework} where children indicated they preferred AI assistants that would offer them support while encouraging them to find solutions independently. The system prompt refers to a set of instructions provided to the LLM (Large Language Model) that dictate how the AI should respond to user inputs. In this instance, the AI assistant's role is to provide a starting point, offer timely advice, give code debugging tips, and encourage a positive learning process that allows for multiple iterations and experimentation.

The AI Copilot’s responses were designed to be brief and child-friendly to avoid complex sentences that might overwhelm young learners. According to the system prompt, the AI Copilot's main objective is to ask questions instead of giving answers, and provide direct help by re-asking the same questions multiple times, only if the user appears to be stuck.

The system prompt was based on feedback from children in our prior study \cite{druga2023aifriendsdesignframework}:
 \begin{quote}
    "You are a helpful assistant for middle school students working on Scratch projects. Keep responses to a single short phrase that's easy for kids to understand and try to first ask kids a question so they can find the answer by themselves. If they don't find the answer and ask the same questions more than 2 times, give them the answer. If someone asks what you can do, say exactly: 'I can help you debug and explain scratch code, give you ideas for making your game more fun, or generate images for your project - just type \\``generate image\'' followed by what you want to see!' For code questions, give one specific tip or ask one guiding question. For project ideas, suggest one fun addition. For code explanations, explain one concept at a time. Remember to keep everything super friendly and encouraging!"
\end{quote}

The AI Copilot is also programmed to respond to the user request, \textit{``What can you do?''} with the phrase, \textit{``I can help you debug and explain scratch code, give you ideas for making your game more fun, or generate images for your project - just type ``generate image'' followed by what you want to see!''}. This allows users to understand the full capabilities of AI Copilot. This approach was based on findings from a prior AI coding assistant co-design study \cite{druga2023aifriendsdesignframework}, where children reported that they appreciated AI assistants that provided positive reinforcement, while also offering guidance and suggestions, rather than directive answers.  

By offering timely advice and using positive and encouraging language, the AI Copilot acts as both a helpful coding assistant and a friendly peer. This approach seeks to balance providing useful assistance and respecting the user's agency in the creative coding process.

While most participants appreciated this design, some participants wanted to personalize the AI Copilot's persona further, for example, preferring the AI to ```always give me 3 ideas so I can select the best''(L. age 12 (Romania)). We plan to allow users to add a system prompt to personalize the AI Copilot in future iterations, allowing users more fine-grained control over the AI’s behavior, therefore tailoring the AI Copilot to match their specific preferences and learning styles.

\subsection{Data Collection and Analysis}
We collected video recordings of 20 study sessions, including two participants who engaged in two sessions each due to initial technical issues, resulting in a rich dataset. Children were encouraged to think aloud \cite{charters2003use} during the collaborative creative coding activities. After the activities, children participated in semi-structured interviews to further explore their experiences and perceptions of coding with the AI Copilot.

Our analysis is primarily qualitative, focusing on the detailed insights gained from the video recordings and interview transcripts. The video data yielded transcripts of 178,105 words. We conducted the sessions in English, Spanish, French, and Romanian. For non-English sessions, the first author transcribed the videos in their original language and subsequently translated them into English to facilitate analysis across all sessions. The first author transcribed all videos, and the transcripts included detailed notes on participants' body language and non-verbal cues to capture a holistic understanding of their interactions.

Our analysis followed an iterative and thematic process. Initially, the authors independently reviewed a subset of the data, focusing on the nuances of collaborative creative coding with the AI Copilot. This involved applying both etic codes (informed by prior literature on creative self-efficacy\cite{tierney2002creative} ) and emic codes (emerging inductively from the data itself) \cite{miles1984drawing,patton1990qualitative}. 

Following this independent review, the authors developed a comprehensive coding frame collaboratively. The first author then coded the majority of the transcripts using the agreed-upon frame, with the second and third author coding a smaller subset to validate the frame. New codes were allowed to emerge throughout this process. When new codes were identified or discrepancies in coding were noted, all authors engaged in discussions to refine code definitions and ensure consistent application. This iterative process resulted in revisions to the coding frame and subsequent re-reading of transcripts to ensure that all data was coded according to the refined structure. The final coding frame, including code definitions and their occurrence frequency, is shown in Table \ref{tab:codebook}. 

The coded data was then synthesized to develop categories, which were conceptualized into broader themes through ongoing discussions among the authors. We employed thematic analysis techniques, drawing upon the principle of saturation \cite{braun2006using}, where the emergence of no new themes towards the end of the analysis indicated that the major themes relevant to our research question had been identified. These themes are presented in the Findings section below.

\begin{table*}[t]
\centering
\begin{tabularx}{\textwidth}{l>{\raggedright\arraybackslash}p{3cm}Xr}
\hline
\textbf{Code} & \textbf{Definition} & \textbf{Example} & \textbf{Count} \\ \hline
\rowcolor[HTML]{EFEFEF}
Conceptual Support & 
\begin{tabular}[c]{@{}l@{}}Helps generate project\\ ideas \end{tabular} & 
\textit{``What about adding secret levels that players can discover?''} & 
12 \\ 

Design Support & 
\begin{tabular}[c]{@{}l@{}}Visual design support\\ and asset generation\end{tabular} & 
\textit{``Generate a crab image... make the maze background less complex''} & 
33 \\ 

\rowcolor[HTML]{EFEFEF}
Code Support & 
\begin{tabular}[c]{@{}l@{}}Provides coding \\ and debugging help\end{tabular} & 
\textit{``Use 'forever loop' with 'move 10 steps' for continuous motion''} & 
46 \\ 

Positive Encouragement & 
\begin{tabular}[c]{@{}l@{}}Offers motivational \\feedback \end{tabular} & 
\textit{``Great idea! Let's try making the ghost character green''} & 
8 \\ 

\rowcolor[HTML]{EFEFEF}
Platform Navigation & 
\begin{tabular}[c]{@{}l@{}}Guides interface\\ exploration\end{tabular} & 
\textit{``Find the 'switch backdrop' block in the Looks category''} & 
12 \\ 

AI Failure & 
\begin{tabular}[c]{@{}l@{}}Unsuccessful or\\ misleading suggestions\end{tabular} & 
\textit{``Generated maze with unwanted portals despite 'simple' request''} & 
20 \\ 

\rowcolor[HTML]{EFEFEF}
Child Agency & 
\begin{tabular}[c]{@{}l@{}}Child chooses not to use\\ the AI and prefers to\\ work independently\end{tabular} & 
\textit{``I don't need the AI for this part, I want to figure it out myself''} & 
9 \\ \hline
\end{tabularx}
\caption{Codebook with code definitions, examples, and occurrence frequencies from 18 participants. Counts reflect total instances across all transcripts.}
\label{tab:codebook}
\end{table*}

\section{Findings}
\begin{figure*}[thpb]
\centering
\includegraphics[width=5.5in]{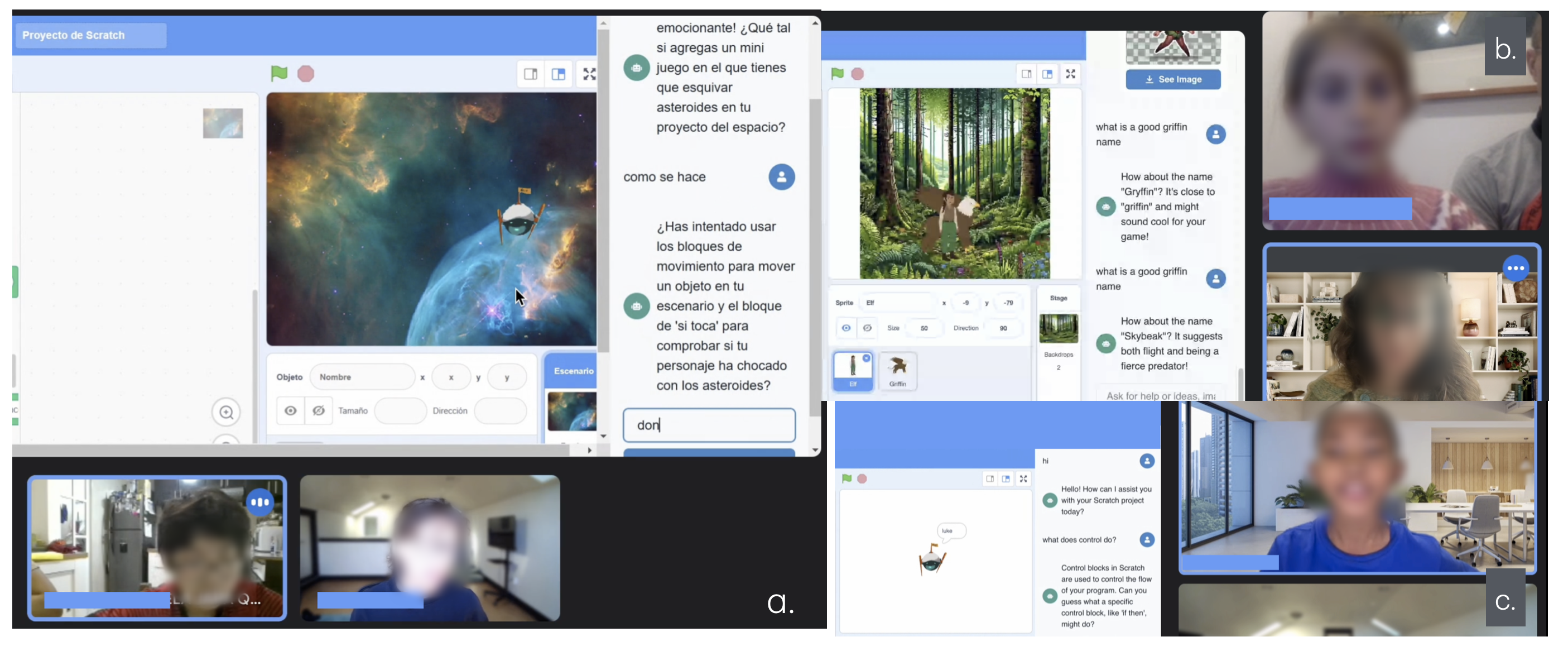}
\caption{Examples of youth-AI Copilot interaction during the study: (a) S., age 11 (Mexico) asking AI for code help for his asteroid game in Spanish , (b) C., age 10 (Canada) brainstorming name ideas for her Griffin character with AI , (c) L., 7 years old (Jamaica) asking AI for intro coding guidance about control blocks.}
\label{fig:fig_child_ai_interaction}
\end{figure*}
Our analysis of the youth creative coding sessions, grounded in our codebook analysis (Table \ref{tab:codebook}) of session transcripts and observational notes, revealed key patterns in how children interacted with and perceived the AI Copilot. The projects undertaken by children varied, including multiplayer games, maze games, storytelling and animation projects, sports games, and collecting games. Across the 18 participants, children used the AI Copilot between \emph{3 and 12 times} per session during the 20-30 minute AI-enhanced coding phase. From our coded data, we identified three overarching themes that capture both the benefits and the design tensions observed:

\subsection{Theme 1: AI-Enhanced Ideation and Asset Creation}
(Related to 'Conceptual Support' and 'Design Support' codes, Table \ref{tab:codebook})
Children used the AI to brainstorm project ideas or generate assets for their projects ('Conceptual Support', 12 instances; 'Design Support', 33 instances).

\begin{figure*}[thpb]
\centering
\includegraphics[width=5.2in]{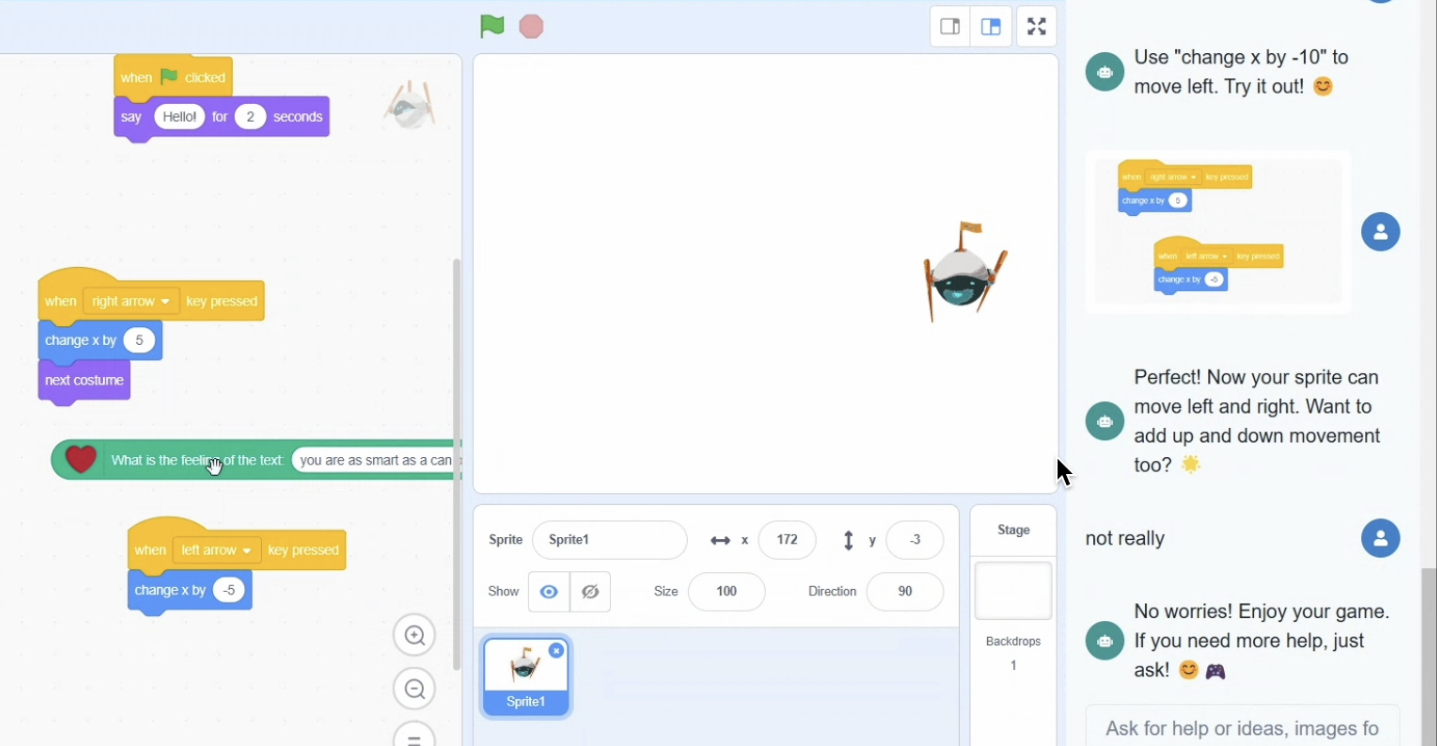}
\caption{Example of ideation support where the child refuses the AI Copilot suggestion.}~\label{agency}
\end{figure*}
\noindent\textbf{Influence on Idea Generation.}
The AI Copilot effectively supported youth in brainstorming project ideas with \emph{13 out of 18} participants explicitly asked the AI Copilot to propose ideas. When participants expressed initial uncertainty or sought inspiration, the AI Copilot offered targeted suggestions and prompts to initiate their creative coding process. For instance, when D. (age 7, USA) was unsure how to expand his game, the AI suggested \textit{``adding secret levels that players can discover,''} sparking the idea of hidden tunnels within his maze.  Similarly, when G.(age 11, USA) was looking for a project idea, the AI Copilot proposed both a \textit{``memory card game''} and a \textit{``virtual pet game''}.  These suggestions served as valuable starting points, helping children overcome blank canvas syndrome and envision potential projects. Beyond initial project ideation, the AI Copilot also assisted with specific game mechanics. When S. (age 11, Mexico) was developing a space-themed project, the AI Copilot suggested adding a mini-game where the player has to \textit{``dodge asteroids''}, enriching the gameplay (Figure \ref{fig:fig_child_ai_interaction}-a):

\begin{quote} \textit{``I was stuck until it said ``secret levels'' – then I imagined hidden tunnels in my maze!''} \ --- D., age 7 (USA)
\end{quote}

While the AI Copilot proved valuable for ideation, its suggestions were not always perfectly aligned with the child's evolving vision:
\begin{quote} 
\textit{``Perfect! Now your sprite can move left and right. Want to add up and down movement too?''} \ --- AI Copilot asking J. a follow-up question after a successful implementation of its suggestion.\\
\textit{``Not really''} \ --- Responded J., age 14 (New Zealand) who proceeded to add a different game feature (Figure \ref{agency}).
\end{quote}

Youth used the AI to also brainstorm character names or ideas for creating new project backgrounds:
\begin{quote} 
\textit{``I was expecting it to give something more unique.''} \ --- C., age 10 (Canada) after they received \textit{Griffin}) as a character name suggestion from the AI (Figure \ref{fig:fig_child_ai_interaction}-b).\\
\textit{`` What do you think will happen if we ask again?''} \ --- asked the researcher.\\
\textit{``Maybe it will give something else(...)Sky Beak is good!''} \ --- C., age 10 (Canada) after trying to get another name idea.
\end{quote}
These examples highlight the importance of iterative prompting and refinement in AI-assisted ideation.

\noindent\textbf{Support with Visual Creation and Design.}

A significant contribution of the AI Copilot was in facilitating visual creation and design. The tool enabled youth to generate visual assets for their Scratch projects, including characters, background images, and props, enhancing the aesthetic appeal and personalization of their creations:

\begin{quote}
\textit{``I want like a zombie. All like zombies all around it''} --- P., age 7 (USA) prompting the AI to generate zombie images for her game.
\end{quote}

 Similarly, W.(age 7, USA), when brainstorming what characters to use in his multiplayer game, considered, \textit{``Um, I don't know. A crab maybe. Yeah, they doesn't have crab (after checking gallery of sprites).''} The lack of pre-existing specific sprites lead him to use the AI-assisted generation of a crab character. The ability to quickly generate diverse visual elements lowered barriers to entry and allowed children to rapidly prototype and visualize their game worlds.

However, the AI-generated images were not always immediately usable and sometimes required refinement.  G. (age 12, Romania/USA) described an initial generated image of a basketball player as \textit{``kind of funny and weird''}, indicating the initial output needed improvement to match his creative vision (Figure \ref{visualsupport}). 

When image generation diverged from expectations, participants developed problem-solving strategies. Furthermore, as highlighted by a participant from Mexico working with monsters images, cultural nuances in design were not always captured by the AI, necessitating manual customization using in-platform drawing tools:

\begin{quote}
\textit{``I asked for a Mexican monster, but it made cartoon dragons. I had to draw the patterns myself.''} \ --- S, age 10 (Mexico).
\textit{``First I said 'simple maze,' then added 'no portals, for kids' – it worked better!''} \\ --- G., age 11 (USA) \end{quote}

\noindent\textbf{Balancing Inspiration vs. Over-Reliance.}
A key tension emerged around whether children might \emph{over-rely} on AI-provided ideas or assets \cite{zhai2024effects}. While children often used the AI’s suggestions as a spark before extending or altering them—mitigating the risk of uncritical acceptance—some older participants (ages 10--12) voiced concern about losing originality. This indicates a potential design tension: \textit{children appreciate the assistance for overcoming creative blocks, yet desire to maintain ownership over the creative direction}.

\subsection{Theme 2: Contextual Debugging and System Navigation}
(Related to 'Code Support' and 'Platform Navigation' codes, Table \ref{tab:codebook})
All participants encountered challenges, whether with coding logic or navigating the platform interface ('Code Support', 46 instances; 'Platform Navigation', 12 instances), prompting them to consult the AI Copilot. In these instances, the AI responded with clarifying questions and tips aimed at guiding the child to find the solution themselves.

\noindent\textbf{Coding and Debugging Support.}
The AI Copilot provided crucial support for coding and debugging, acting as a readily available source of guidance.  It offered specific coding instructions and troubleshooting assistance across various coding challenges. For example, when facing difficulty resizing a sprite, L., age 7 (Jamaica) was prompted by the AI:

\begin{quote}
\textit{``How can I make my character smaller?''} ---L., age 7 (Jamaica)
\textit{``Have you tried changing the size of your sprite by using the set size block?''} --- AI Copilot responding to W., age 7 (USA).
\end{quote}

 Similarly, when L.(age 7, Singapore) needed to implement a repeating action, the AI suggested:\textit{``Have you tried using a forever loop with a move block inside?''}  For more complex coding concepts, like making a sprite bounce, the AI provided step-by-step explanations: \textit{``To make a Sprite bounce back, you need to change the direction of the Sprite by 180 degrees. When it it bounce, it touches the edge.''}  This support extended to explaining how to achieve specific actions like \textit{``How to make a sprite disappear,''} \textit{``How to move a sprite,''} or \textit{``How to make a sprite bounce,''}  effectively demystifying coding tasks for novice users.

However, the AI Copilot's code support was not without limitations.  At times, it struggled with nuanced or ambiguous queries, providing inaccurate, overly general, or incomplete guidance.  For instance, when W.(age 7, USA) asked to make a crab smaller, the AI initially misunderstood the question and did not provide a useful answer, prompting the researcher to clarify: \textit{``It did not understand your question. What did it say? Maybe you can ask to make the crab smaller in Scratch? ''}  

When coding answers were not helpful for their particular project needs, older participants articulated strategic help-seeking:
\begin{quote}
\textit{“I’ll ask for the loop structure but write the variables.”} — K., 10 (Romania/USA).
\end{quote}

Several participants (10 out of 18) expressed the desire for the AI Copilot to be able to observe what they do on the screen and nudge them if they are doing something wrong or better understand their questions: 
\begin{quote}
\textit{``You know what, I need this also. I need this thing there. It should be able to fit through the gap.''} --- W., age 7 (USA)
\end{quote}

These findings highlight the need for more robust intent recognition and context awareness when designing AI coding assistants for children.

\noindent\textbf{Support with Platform Navigation \& Onboarding}
Beyond coding-specific assistance, the AI Copilot also served as a valuable navigation aid within the Scratch environment. For children new to the platform, navigating the block palettes and understanding the interface presented an initial hurdle (Figure \ref{fig:fig_child_ai_interaction}-c). The AI Copilot helped overcome this by guiding users to specific block categories and functionalities. When L. (age 7, Singapore) needed to change the background, the AI directed him: 

\begin{quote}
\textit{``Did you see that block that says switch background too in the looks category''} --- AI Copilot to L., age 7 (Singapore).
\end{quote}

Additionally the AI Copilot supported youth to discover how to save their project, how to edit the background of characters, how to search for sounds or load a local file to the background. While helpful, platform navigation remained a learning process. Children were novices to Scratch, and even with AI guidance, some instructions required further scaffolding from the researcher.  For example, the first author suggested using copy-paste command for text of prior questions to AI: \textit{``You can  duplicate your text from above and edit it ''} to T. (age 7, Romania), assuming a certain level of digital literacy that might not be universal among young learners. The AI Copilot acted as a supplementary tool, easing the initial navigation learning curve, but not fully replacing the need for some direct scaffolding, especially for younger or less digitally experienced users.

\noindent\textbf{Researcher Facilitation Frequency.}
On average, the AI successfully answered about \emph{70\% of queries} (estimate across sessions). In the remaining \emph{30\%} ('AI Failure' code, 20 instances total), the first author had to intervene – occurring \emph{two to three times per session on average} – to clarify ambiguous AI feedback, correct incomplete instructions, or help when the AI misunderstood the child’s intent. This scaffolding approach provided support while still giving the AI the first chance to help.

\subsection{Theme 3: Preserving Child Agency and Addressing Potential Negative Effects}
(Related to 'Child Agency' and 'AI Failure' codes, Table \ref{tab:codebook})
In some cases, children chose to adapt or not use AI suggestions('Child Agency', 9 instances) and we also identified several examples when the AI did not provide useful suggestions or it was misleading ('AI Failure', 20 instances).

\noindent\textbf{Adapting or Rejecting AI Suggestions.}
Despite appreciation for the AI’s assistance, participants across the entire age range (7--12) demonstrated strong ownership over their projects ('Child Agency' code explicitly noted 9 times, but observed more broadly). Children frequently evaluated AI suggestions, adapting or rejecting them to fit their vision. In \emph{9 of 18} cases, children explicitly declined an AI’s suggestion:

\begin{figure*}
\centering
\includegraphics[width=5.2in]{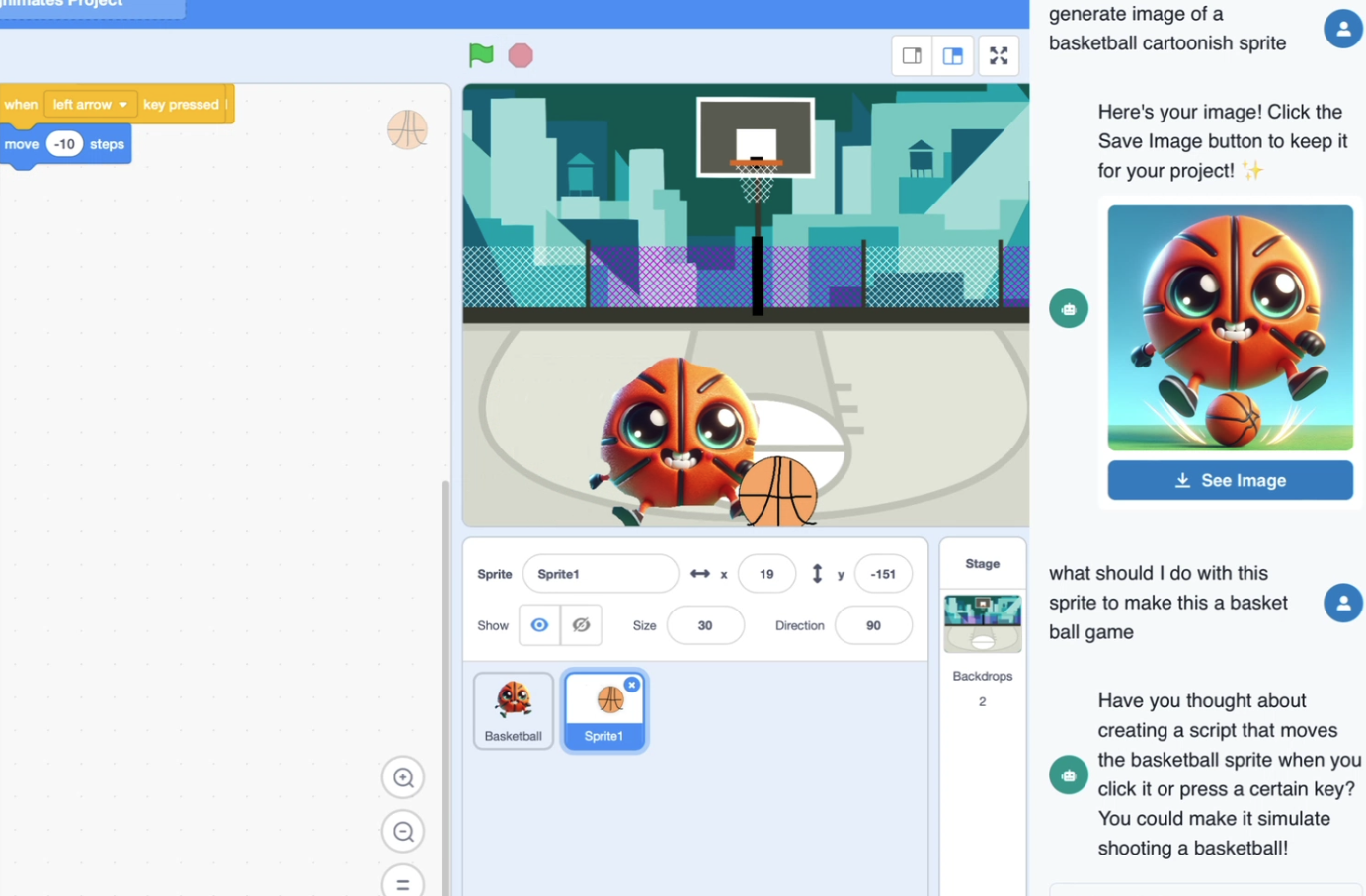}
\caption{Example of visual creation support for G., age 12 (Romania) who wanted a custom basketball player character.}
~\label{visualsupport}
\end{figure*}
Participants highly valued the AI Copilot's assistance, yet they consistently emphasized the importance of maintaining their own creative ownership and control throughout the coding process:

\begin{quote}
\textit{``It's like a teammate who knows coding tricks, but I'm the captain.''} \ --- K., age 10 (Romania/USA)
\end{quote}

Older children, like K. above, articulated a nuanced perspective on human-AI collaboration. This metaphor of being the captain highlights the child's desire to remain the creative director, leveraging the AI as a helpful assistant rather than a co-creator.

Younger children demonstrated their agency through more action-oriented behaviors.  For instance, W. (age 7, USA), despite receiving AI-generated code suggestions, would often physically drag the suggested blocks aside to manually experiment and understand their function before integrating them into his project. This hands-on exploration reflects a desire to deeply learn and internalize the coding concepts, rather than simply accepting AI-provided solutions.  Crucially, children sometimes actively chose to diverge from AI suggestions to pursue their own creative visions, even if it meant deviating from a potentially more technically ``optimal'' path:

\begin{quote}
\textit{``The bouncing code worked, but I wanted my crab to flip upside-down!''} \ --- W., age 7 (USA)
\end{quote}

This deliberate choice to prioritize personal creative expression over AI-guided efficiency underscores the importance of designing AI tools that enable, rather than replace, child agency in creative coding.  

Youth manifestation of agency when interacting with the AI Copilot revealed three patterns: direct rejection of AI suggestions, adaptive integration when children would modifying the AI suggestions, and preemptive control:
\begin{quote}
\textit{``No, that’s not my idea!''} \ --- N., age 8 (USA), reacting to AI Copilot coding suggestion to add a timer to his game.\\
\textit{``Wait, let me try first.!''} \ --- G., age 12 (Romania), refusing the researcher's suggestion to ask the AI for help when trying to debug his code.
\end{quote}
Children actively governed AI use, with 9 out 18 rejecting at least one suggestion. 

\noindent\textbf{Potential Over-Reliance and Navigating AI Errors.}
Some children initially appeared to assume AI suggestions were always correct. However, encountering instances where the AI provided incorrect code, misunderstood context, generated unusable images, or gave incomplete responses surfaced a tension: \emph{AI can accelerate progress, but also derail novices who might overtrust it}. Examples of AI failure included:
\begin{enumerate}
\item

\textbf{Misunderstanding Intent}: The AI Copilot Tool often struggled with understanding the user’s underlying intention, especially with vague questions:
\begin{quote}
\textit{``Make it bigger?''} --- T., age 9 (Israel), referring to her onscreen character.  
\end{quote} 
\item \textbf{Generating unusable or unexpected images:} At times, the AI  generated images that did not align with the user’s expectations, such as creating a maze with portals, even after the child tried to get the maze to be less complex, or generating an image of a character that looked like a Minecraft world, instead of a Roblox avatar:
\begin{quote} 
\textit{``Generate image of Roblox''} --- L., age 7 (Jamaica)
\end{quote}
\item \textbf{Incomplete responses:} Although \textbf{AI Copilot Tool} provided helpful solutions, it sometimes gave incomplete instructions on how to achieve that specific step.
\begin{quote}
\textit{``I don't know where the ``hide'' block is."}--- T., age 7 (Romania)
\end{quote}
\end{enumerate}

Instances where the AI Copilot did not meet expectations, or demonstrably ``failed,'' were not simply negative outcomes but often transformed into valuable learning opportunities.  These breakdowns prompted productive troubleshooting and deeper engagement with the coding process.  When AI image generation produced unsatisfactory results, participants actively refined their prompts, engaging in iterative problem-solving. C (age 10, Canada) recounted her prompt refinement process for an elf character generation: \textit{`` I asked for an elf then added 'full body and no background.''}  This trial-and-error approach to prompt engineering not only improved the AI's output but also fostered metacognitive skills in articulating intent and understanding the AI's limitations.

Furthermore, AI failures sometimes surfaced fundamental programming concepts in a more tangible way.  When K. (age 10, Romania/USA) encountered difficulties positioning AI-generated sprites, it led to a moment of insight about coordinate systems:

\begin{quote} 
\textit{Why does X=0 put it in the middle? That's confusing... Oh, wait! Maybe like a number line!''} \ --- K., age 10 (Romania/USA)
\end{quote}

This ``aha'' moment, triggered by an AI-related challenge, demonstrates how breakdowns can also create ``teachable moments,'' prompting deeper conceptual understanding.

\subsection{Final Interview Reflections: AI Copilot Perceptions and Prior AI Exposure}

Following the coding sessions, semi-structured interviews provided valuable insights into children's perceptions of the AI Copilot, their AI literacy, and their broader experiences with AI in their lives.  Overall, children expressed positive perceptions of the AI Copilot, consistently describing it as \textit{``helpful,''} \textit{``good,''} \textit{``cool,''} and \textit{``fun.''}  This positive sentiment was often linked to the AI's ability to provide coding assistance  when needed, acting as a readily available guide to overcome coding hurdles and generate creative assets:

\begin{quote}
\textit{``Well I feel like the AI really helped.''} --- P., age 7 (USA) when asked by the researcher to describe the interaction with AI. 
\end{quote}

The interviews also revealed diverse levels of AI exposure and real-world AI experiences among the participants.  Most children had heard of AI, and their definitions often centered on AI as a \textit{``human made''} technology capable of \textit{``thinking for itself''} but ultimately \textit{``programmed kind of,''} as described by W. (age 7, USA).  L. (age 7, Singapore) succinctly defined AI as \textit{``I think AI is a robot,''} reflecting a common association of AI with embodied intelligent agents.  Children from North America demonstrated greater familiarity with real-world AI applications. N (age 8, USA) mentioned using AI \textit{``sometimes with my dad at home''} and being aware of \textit{``Chat GPT''}.  In contrast, children from other regions often had more limited direct experience, with L (age 7, Singapore) stating his AI experience was \textit{``Only for Google,''} and others participants like I. (age 7, Spain) and L. (age 7, Jamaica) having no prior AI interaction of any kind.  This variability in AI exposure levels highlights the importance of contextually appropriate AI education and design considerations for diverse youth populations.

\section{Discussion}

Our work asked, \textit{How might we support children engaging in collaborative creative coding with an AI Copilot?} Our study of Cognimates Scratch Copilot revealed that such a tool can provide multifaceted support across the creative coding process, but its integration requires careful consideration of pedagogical goals, user agency, and potential challenges. We discuss our findings in relation to our three core themes: AI-enhanced ideation and creation, contextual debugging and navigation, and preserving child agency while addressing potential negative effects.

\textbf{AI's Role in Ideation, Creation, and Potential Over-reliance}: The AI Copilot proved particularly effective in supporting the initial, often challenging, phases of creative work (Theme 1). By offering targeted suggestions for project ideas, game mechanics, and character concepts, it helped youth overcome the ``blank canvas'' problem and translate abstract ideas into starting points, aligning with prior work highlighting the need for scaffolding in creative tasks \cite{Frich2019-zj, Kim2017-mosaic}. The integrated image generation was highly engaging, lowering barriers to visual design and allowing rapid prototyping. However, this support surfaced a key tension: the potential for over-reliance. While participants often adapted or iterated on AI suggestions, preserving their vision, the ease of generation necessitates designs that encourage customization and critical evaluation, rather than passive acceptance. The need for prompt refinement and manual editing when AI outputs were unsatisfactory highlighted both a limitation and a learning opportunity, fostering nascent prompt engineering skills.

\textbf{Contextual Support, Scaffolding, and AI Limitations:} Cognimates Scratch Copilot served as a readily available resource for debugging and navigating the coding environment (Theme 2), offering timely suggestions and explanations that could demystify coding concepts for novices. These results align with previous research highlighting the benefits of collaboration, scaffolding, and exploration in creative coding experiences for children \cite{resnick2007all,moralesnavarro2024investigatingyouthseverydayunderstanding}. By actively involving AI in the creative process, the AI Copilot Tool allowed youth to bridge the gap between their ideas and the actual implementation of code.  This is particularly important for young learners, as translating abstract ideas into concrete code can be a significant hurdle \cite{Ko2004}. As Ko et al. point out, novice programmers face learning barriers related to syntax, programming concepts, and debugging, and these are even harder to overcome for children, given less prior knowledge and less developed metacognition. Furthermore, AI tools can serve as both a starting point for novices and provide support to more advanced users, by offering more complex recommendations and by guiding them through more intricate programming tasks.

However, there were several limitations to AI Copilot support, showing a need for future research to consider nuanced challenges, diverse user needs and preferences, and the impact of AI in creative coding. Participants wanted the AI Copilot to be more precise, to provide options instead of a single solution, to understand nuances of language, and to understand the context of the game they were trying to build.  They also expressed a desire for the AI to tailor its support to their specific needs and prior experiences, echoing findings in personalized learning research \cite{tetzlaff2021developing}.

\textbf{Agency, Ethical Considerations, and Learning from Failure}: Perhaps the most critical finding relates to the preservation of child agency (Theme 3). Participants consistently demonstrated a desire to remain ``the captain'' of their creative process, actively rejecting, adapting, or preempting AI suggestions to maintain control. This aligns with constructionist learning principles emphasizing learner control and exploration \cite{papert1980mindstorms, resnick2013designing} and resonates with ethical considerations for AI in education that prioritize learner empowerment \cite{newman2024want, morales2024unpacking}. Children wanted AI to provide support for problem-solving, but not solve the problem for them. They also emphasized the importance of transparency to understand how their code works and to make their own informed creative decisions, highlighting the importance of balancing AI support with user agency. This desire for agency and transparency is consistent with ethical considerations for AI in education, emphasizing the need for AI to be interpretable and to empower learners rather than replace them \cite{newman2024want, morales2024unpacking}. Children's explicit concerns about ``getting lazy'' or losing originality highlight their own awareness of the potential downsides of over-reliance. Furthermore, AI failures were not merely obstacles but often transformed into productive learning moments, prompting deeper engagement with debugging, prompt refinement, and even conceptual breakthroughs (like understanding coordinate systems). This suggests that encountering and overcoming AI limitations can foster resilience and critical thinking, turning potential negative effects into pedagogical opportunities. Designing for transparency and encouraging critical evaluation of AI outputs seem crucial for mitigating risks like the uncritical acceptance of potentially flawed suggestions.

The study sessions also showed children had a clear understanding of AI as a tool for learning, while expressing the need to not rely too heavily on the tool, to maintain agency and control over their learning experience. These results build upon prior work about how children learn with and about AI \cite{druga2022family,long2022family, morales2024unpacking}. Children recognized the copilot as a helpful learning resource but also wanted to develop their own problem-solving abilities.

Furthermore, these findings were reinforced by parental feedback shared via emails after the sessions. For example one parent noted that their child ``totally rejected the idea of becoming too dependent on it [referring to the AI Copilot] and perhaps not remembering or learning how to do it on her own. She said that she would just use it to learn''. Another parent also mentioned that participation in the study for their child ``has stimulated her desire to do some coding at home in Scratch’’, suggesting the potential of AI Copilots to have a positive influence in the learning journey of youth by sparking interest and motivation.

\textbf{Need for multimodal inputs and outputs:} The system, at times, did not always meet the user’s creative vision, leading sometimes to children having to make use of drawing tools in the interface or re-generate images to better fit their project needs. This indicated that image creation capabilities should be fully integrated with the coding process, and not be an afterthought. Children also expressed a desire for voice-based interaction, although they still said they preferred the text-based chat, showing a preference for modalities to be flexible in AI interfaces. This aligns with research on multimodal learning environments, suggesting that offering varied input and output modalities can enhance engagement and cater to diverse preferences \cite{mayer2005cognitive}.

\textbf{Designing culturally responsive tools:} Furthermore, our study revealed diverse responses based on experience levels, which highlighted the need for better AI personalization. Children who had more previous experience with block-based coding and who had been exposed to AI in their day to day, had different expectations from the AI assistant, requiring more advanced features (i.e. custom system prompt) that allowed them more control and customization.  Moreover, exposure to AI varied among participants depending on geographical location, which affected their level of fluency in interacting with the AI Copilot. Children from North America had more familiarity with various AI technologies compared to children from the other regions. These findings underscore the importance of culturally responsive computing education and the need to design AI tools that are adaptable and inclusive, supporting a wide range of learners with varying prior knowledge and experiences \cite{solyst2022insights}.

Overall, our findings suggest that AI copilots like Cognimates Scratch Copilot hold significant potential to enhance youth creative coding by providing timely support for ideation, implementation, and debugging. However, realizing this potential requires designs that actively prioritize child agency, offer flexible and context-aware support, balance guidance with challenge, and frame the AI as a collaborative partner rather than an infallible authority.

\subsection{Design Guidelines}
Based on the findings of our study we propose that future designs of AI Copilots for youth should take into consideration the following design guidelines:
\begin{itemize}
  \item  \textbf{Prioritize user agency:} Young learners value having control and not seeing AI do all the work. The AI should serve as a helper, not a replacement of their own creative work.
  \item  \textbf{Balance support and challenge:} Provide AI support but with enough open-endedness for exploration and problem-solving, to provide a balance between challenge and assistance.
  \item \textbf{AI as a Motivator and Starting Point:} AI can help kids getting started, but also allow them to learn faster by providing a starting point and scaffolding their learning process.
 \item \textbf{Allow Flexibility and Customization:} AI Copilot platforms should allow for multiple ways for a user to learn and interact with the tool and choose if they want to use voice or text for input, if they want to allow the AI to move blocks or not, and allow youth to customize the type of responses they get via an editable system prompt (i.e. ``I want it to always give me 3 ideas’').
 \item \textbf{Design AI tools to support visual creativity in-situ:} The image generation aspect of the platform is highly engaging, and needs to be more integrated with the coding platform context so that generated images can be directly added as platform characters or backgrounds and would be the visual style that youth expects from the platform (i.e. more like a cartoon rather than a realistic image, without background and full body).
  \item \textbf{Support Multimodal AI capabilities:} AI should be able to generate both text and sounds, and also to be able to understand voice input, while allowing users to select their preferred way of interaction with the tool.
  \item \textbf{AI should acknowledge the user's preferences and limitations:} The system should allow users to personalize its prompt and responses, to address diverse youth needs and preferences.
\end{itemize}

\subsection{Limitations and Future Work}
Our work had certain limitations. First, we conducted a small sample study with 18 participants in a relatively short time frame. Future work should examine these effects across more users, different demographics, and over longer periods of time. Second, the study did not focus on code sharing and collaboration, but only on individual sessions with the AI Copilot. More work is needed on how to support collaboration and co-creation through AI-driven tools. Finally, the study focused on a single tool (a visual programming environment similar to Scratch) and may not generalize well to other creative coding environments. Future work should also focus on how to adapt our design guidelines for other platforms with different features and specific affordances.

While not a primary focus of our coded analysis, our observations suggested potential developmental differences in how children interacted with the AI Copilot. Younger participants (ages 7-9) appeared to engage with the AI more conversationally, appreciating its immediate responses and perceiving it as a helpful companion. They also expressed high satisfaction with the AI’s basic functionalities.  In contrast, older participants (ages 10-12) exhibited greater independence, utilizing the AI for more complex problem-solving and advanced creative explorations. These preliminary observations suggest that future research should systematically investigate developmental nuances in AI-assisted creative coding, exploring how AI Copilots can be designed to adapt to varying developmental stages, expectations, and prior experiences. Further studies with larger, age-stratified samples are needed to validate these initial observations and develop age-appropriate design guidelines for AI-powered learning tools for youth.

Some of our study participants expressed awareness of some of the challenges of AI, in particular about the risks of misinformation or the potential for it to be used to produce fake content demonstrating a nascent critical AI literacy interest even in young children. Future studies should further investigate childrens' perceptions of AI risks in the context of code assistance for youth.

\section{Conclusion}
This paper presented Cognimates Scratch Copilot, an AI-powered assistant designed to support youth creative coding within a visual block-based environment. Our exploratory study with 18 international children demonstrated that such a tool can effectively scaffold the creative coding process by aiding ideation, providing coding and debugging assistance, facilitating asset creation, and helping with platform navigation. However, our findings also highlight the critical importance of designing these tools to preserve child agency, foster critical engagement, and manage potential challenges like over-reliance or navigating AI errors.

The results underscore that AI copilots, when designed thoughtfully according to principles like prioritizing user control, balancing support with challenge, and promoting transparency, have the potential to empower youth, enhance their creative self-efficacy, and deepen their engagement with computational thinking. They can act as valuable partners in the learning journey, offering guidance while encouraging creative independence. Future work must continue to refine these tools, focusing on contextual awareness, personalization, collaborative support, and fostering critical AI literacy, to ensure that AI truly serves to augment, rather than diminish, the creative and learning processes of the next generation of creative coders.

\begin{acks}
 This material is based upon work supported by the National Science Foundation under Grant No. 1539179, 1703304, 1836813, 2031265, 2100296, 2122950, 2137834, 2137312, 2318257, 2417014 and unrestricted gifts from Microsoft, Adobe, and Google.
\end{acks}

\bibliographystyle{ACM-Reference-Format}
\bibliography{uwthesis,refs_tc,refs_ge,refs_ai,refs_idc21,refs_ge2,refs_thesis,refs_citizen,refs}
\end{document}